\documentclass[english]{sbrt}
\IEEEoverridecommandlockouts
\usepackage{amsmath,amsfonts}
\usepackage[ruled,vlined]{algorithm2e}
\usepackage{array}
\usepackage[caption=false,font=normalsize,labelfont=sf,textfont=sf]{subfig}
\usepackage{textcomp}
\usepackage{stfloats}
\usepackage{url}
\usepackage{verbatim}
\usepackage{graphicx}
\usepackage{cite}
\usepackage{xcolor}
\usepackage[nolist,nohyperlinks]{acronym} 
\usepackage{amsmath,amssymb,amsfonts}
\usepackage{algorithmic}
\usepackage{graphicx}
\usepackage{textcomp}
\usepackage{caption}
\usepackage{subcaption}
\usepackage{xspace} 
\def\BibTeX{{\rm B\kern-.05em{\sc i\kern-.025em b}\kern-.08em
    T\kern-.1667em\lower.7ex\hbox{E}\kern-.125emX}}

\begin{document}
\begin{acronym}
	\acro{IRS}{Intelligent reflecting surface}
	\acro{RIS}{reconfigurable intelligent surface}
	\acro{irs}{intelligent reflecting surface}
	\acro{PARAFAC}{parallel factor}
	\acro{TALS}{trilinear alternating least squares}
	\acro{BALS}{bilinear alternating least squares}
	\acro{DF}{decode-and-forward}
	\acro{AF}{amplify-and-forward}
	\acro{CE}{channel estimation}
	\acro{RF}{radio-frequency}
	\acro{THz}{Terahertz communication}
	\acro{EVD}{eigenvalue decomposition}
	\acro{CRB}{Cramér-Rao lower bound}
	\acro{CSI}{channel state information}
	\acro{BS}{base station}
	\acro{MIMO}{multiple-input multiple-output}
	\acro{NMSE}{normalized mean squared error}
	\acro{2G}{Second Generation}
	\acro{3G}{3$^\text{rd}$~Generation}
	\acro{3GPP}{3$^\text{rd}$~Generation Partnership Project}
	\acro{4G}{4$^\text{th}$~Generation}
	\acro{5G}{5$^\text{th}$~Generation}
	\acro{6G}{6$^\text{th}$~generation}
	\acro{E-TALS}{\textit{enhanced} TALS}
	\acro{UT}{user terminal}
	\acro{UTs}{users terminal}
	\acro{LS}{least squares}
	\acro{KRF}{Khatri-Rao factorization}
	\acro{KF}{Kronecker factorization}
	\acro{MU-MIMO}{multi-user multiple-input multiple-output}
	\acro{MU-MISO}{multi-user multiple-input single-output}
	\acro{MU}{multi-user}
	\acro{SER}{symbol error rate}
	\acro{SNR}{signal-to-noise ratio}
	\acro{SVD}{singular value decomposition}
\end{acronym}
\title{Joint Channel and Symbol Estimation for Communication Systems with Movable Antennas}

\author{Josué V. de Araújo, Jose Carlos da Silva Filho, Gilderlan T. de Araújo\\ Paulo R. B. Gomes and André L. F. de Almeida
\thanks{Josué V. de Araújo, Jose Carlos da Silva Filho, and André L. F. de Almeida are with the Teleinformatics Department, Federal University of Ceará, Fortaleza-CE, e-mail: \{josue.vas,jcarlos.sf\}@alu.ufc.br; andre@gtel.ufc.br.}%
\thanks{Gilderlan T. de Araújo and Paulo. R. B. Gomes are with the Federal Institute of Ceará, e-mails: \{gilerlan.tavares,gomes.paulo\}@ifce.edu.br.}
\thanks{The authors thank the partial support of FUNCAP under grant ITR-0214-00041.01.00/23, and the National Institute of Science and Technology (INCT-Signals) sponsored by Brazil's National Council for Scientific and Technological Development (CNPq) under grant 406517/2022-3. This work is also supported by CNPq under grant 312491/2020-4 and by Ericsson Research, Sweden, and Ericsson Innovation Center, Brazil, under UFC.52
Technical Cooperation Contract Ericsson/UFC.}
}

\maketitle

\markboth{XLIII BRAZILIAN SYMPOSIUM ON TELECOMMUNICATIONS AND SIGNAL PROCESSING - SBrT 2025, SEPTEMBER 29TH TO OCTOBER 2ND, NATAL, RN}{}

\begin{abstract}
Communication systems aided by movable antennas have been the subject of recent research due to their potentially increased spatial degrees
of freedom offered by optimizing the antenna positioning at the transmitter and/or receiver. In this context, a topic that deserves attention is channel
estimation. Conventional methods reported recently rely on pilot-assisted strategies to estimate the channel coefficients. In this work, we address
the joint channel and symbol estimation problem for an uplink \acl{MU} communication system, where the \acl{BS} is equipped with
a movable antenna array. A semi-blind receiver based on the PARAFAC2 model is formulated to exploit the tensor decomposition structure for the received signals, from which channel and symbol estimates can be jointly obtained via an alternating estimation algorithm. Compared with reference schemes, our preliminary numerical
simulations yield remarkable results for the proposed method.
\end{abstract}


\renewcommand\baselinestretch{.85}

\section{Introduction}

In recent years, movable or fluid antennas have emerged as a promising research direction in wireless communications, particularly within the context of future 6G networks and integrated sensing and communication systems \cite{yong_2024}. These antennas, which can dynamically reposition themselves within a predefined spatial region, offer the ability to adapt to changing channel conditions and maximize spatial diversity. According to \cite{historico_movable}, movable and fluid antennas are often used interchangeably, as they share the same underlying principle of flexible antenna positioning, regardless of their implementation differences. Both architectures aim to enhance spatial degrees of freedom (DoFs) at the transceiver ends, enabling more efficient spectrum utilization and improved link reliability \cite{Ruoyu_2024,Zhang_2025}.

One of the key challenges in unlocking the full potential of these reconfigurable systems is the development of effective channel estimation techniques \cite{Lipeng_2023}. In \ac{MU} communication systems, where multiple users transmit simultaneously to a central \ac{BS}, accurate channel state information (CSI) becomes essential for tasks such as detection, decoding, and resource allocation. However, using movable antennas introduces time-varying spatial configurations, increasing estimation processing complexity. Existing works have addressed this problem \cite{Ruoyu_2024,Zhang_2025,Xiao_2024,Ma_2023_CE,Ioannis_023_CE}.  A channel estimation method for a movable antenna array on both sides was proposed in \cite{Ruoyu_2024}. The authors of \cite{Zhang_2025} proposed the successive Bayesian reconstructor-based channel estimation method for fluid antenna. In \cite{Xiao_2024}, a general channel estimation framework for movable antenna systems was developed. A compressed sensing approach was proposed by \cite{Ma_2023_CE}, while a MMSE solution in a large cellular network was derived in \cite{Ioannis_023_CE}. This solution relies on pilot-based techniques, which, despite their practicality, often suffer from large training overhead and reduced spectral efficiency, particularly in scenarios with limited coherence time or many users.

To mitigate these limitations, we propose a tensor-based semi-blind receiver for joint channel and symbol estimation in uplink \ac{MU} systems with movable antennas \ac{BS}. Unlike conventional estimators, our approach exploits the inherent multilinear structure of the received signal data, which naturally exhibits a third-order tensor format. We show that this tensor follows a PARAFAC2 model. Tensor signal processing has effectively solved various wireless communication problems \cite{Ximenes_TSP_2014,Ximenes_TSP_2014,Favier_TSP_2014}, including MIMO systems, millimeter-wave communications \cite{DEALMEIDA2007337, 4524041, 6897995, 7152972}, and more recently, reconfigurable intelligent surfaces (RIS) \cite{Gil_SAM,Chen2021,Gil_JTSP,Ruoyu_2024}. Unlike the conventional PARAFAC model, the PARAFAC2 is still not fully exploited in wireless communication. To cite some of this work \cite{Sorensen_2009} to model the received signal as a PARAFAC2 model in an orthogonal space-time block codes (OSTBC) MIMO system. Still considering an OSTBC system, in particular using an Alamouti coding, \cite{Gil_PARAFAC2_2017} started from the received signal, which has been modeled as the PARAFAC2 model, and derived the PARATUCK2 model after algebraic manipulation.


Taking advantage of the PARAFAC2 structure, we formulate a semi-blind receiver capable of extracting joint estimates of the channel and symbol vectors associated with each user via a \ac{BALS} method. This approach avoids or reduces the pilot sequence requirement, improving the spectral efficiency. Our numerical experiments validate the proposed methodology, showing that it achieves competitive performance in normalized mean square error (NMSE) and bit error rate (BER) compared to existing reference techniques. Furthermore, the method remains robust even in scenarios with high spatial compression (i.e., when the number of RF chains is significantly smaller than the number of ports), underscoring its practical viability.

To summarize, the main contribution of this paper is that we consider movable antennas on a two-time scale in wireless communication where a \ac{MU} single antenna communicates with a movable antenna \ac{BS} in the uplink side. On the receiver side, a switching matrix selects the port that must be connected to the RF chain, while on the transmitter side, the data vector is used as Khatri-Rao coding that spreads the information over time. The switching matrix and the coding remain constant within a block transmission and vary block-to-block. The algebraic structure of the received signal follows a PARAFAC2 model. By exploiting such a structure, we formulate a joint semi-blind receiver with an \ac{BALS} solution.

\vspace{2ex}
\noindent \textit{Notation and properties}: Vectors are denoted by boldface lowercase letters ($\mathbf{a})$. Matrices with boldface capital letters ($\mathbf{A})$, while tensors are symbolized by calligraphic letters $(\mathcal{A})$. Transpose and pseudo-inverse of a matrix $\mathbf{A}$ are denoted as $\mathbf{A}^{\text{T}}$ and $\mathbf{A}^\dagger$, respectively. $\mathbf{I}_{N}$ denotes a $N \times N$ identity matrix. $D_i(\mathbf{A})$ is a diagonal matrix holding the $i$-th row of $\mathbf{A}$ on its main diagonal. The operator $\textrm{diag}(\mathbf{a})$ forms a diagonal matrix out of its vector argument. $\|\cdot\|_{\text{F}}$ denotes the Frobenius norm, while the symbol $\otimes$ represents the Kronecker product. The operator $\textrm{vec}(\cdot)$ vectorizes an $I \times J$ matrix argument, while $\textrm{unvec}_{I \times J}(\cdot)$ does the opposite operation. Moreover, $\mathbf{A}_{(i,j)}$, $\mathbf{A}_{(i,:)}$ and $\mathbf{A}_{(:,j)}$ denote the $(i,j)$-th entry, the $i$-th row, and the $j$-th column of matrix $\mathbf{A}$, respectively. In this paper, we make use of the following identities
\begin{equation}
\textrm{vec}(\mathbf{ABC}) = (\mathbf{C}^{\textrm{T}} \otimes \mathbf{A})\textrm{vec}(\mathbf{B}),
\label{Eq:Propertie Vec General}
\end{equation}
\begin{equation}
(\mathbf{A} \otimes \mathbf{B})(\mathbf{C} \otimes \mathbf{D}) = (\mathbf{A}\mathbf{C}) \otimes (\mathbf{B}\mathbf{D}).
\label{Eq:Propertie Mixed}
\end{equation}

\section{PARAFAC2 Overview}


The PARAFAC2 \cite{parafac2} can be seen as a tensor decomposition that relaxes some constraints of the conventional \textit{parallel factors} (PARAFAC) decomposition \cite{Harshman70}. More specifically, while in the PARAFAC model two factor matrices remain constant along two dimensions (forming a regular tensor), the PARAFAC2 decomposition allows dealing with a set of variant matrices (along the first dimension) with the same number of columns but different row sizes. To make the understanding more straightforward, the PARAFAC2 decomposition of a third-order tensor can be represented in terms of its frontal slice notation according to the following expression





\begin{equation}
    \mathbf{X}_p = \mathbf{A}_pD_p(\mathbf{C}) \mathbf{B}^\text{T}, \,\,\, p=1, \ldots, P,
    \label{EQ: generic PARAFAC model}
\end{equation}
where the 3-mode dependent factor matrix is $\mathbf{A}_p \in \mathbb{C}^{I_p \times R}$, while $\mathbf{B} \in \mathbb{C}^{I_2 \times R}$ and $\mathbf{C} \in \mathbb{C}^{I_3 \times R}$ are fixed factor matrices. Generally, the PARAFAC2 decomposition presented in (\ref{EQ: generic PARAFAC model}) is not unique \cite{Kolda_2009}. However, under additional constraints, uniqueness can be reached by imposing $\mathbf{A}_p^{\text{T}}\mathbf{A}_p = \boldsymbol{\Phi}$, $p = 1, \ldots, P$. This means $\mathbf{A}_p = \mathbf{M}_p\mathbf{N}$, where $\mathbf{M}_p\mathbf{M}_p = \mathbf{I}_{I_p}$ \cite{Unicidade_PARAFAC2}.



\begin{figure}[!t]
    \centering
    \begin{minipage}[b]{0.47\textwidth}
        \centering
        \includegraphics[width=\textwidth]{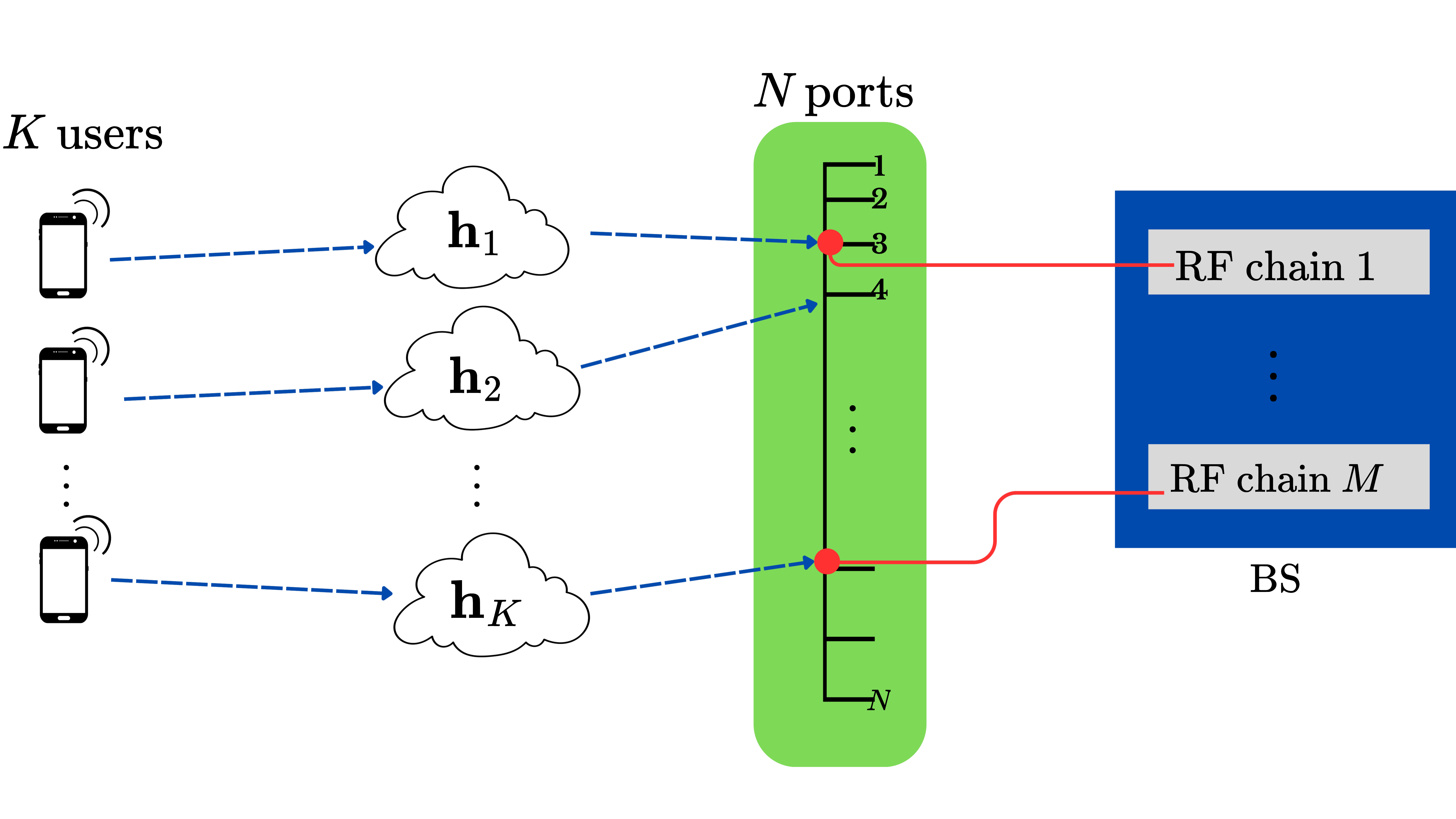}
        \caption{\small{Movable-antenna based uplink \ac{MU} scenario.}}
        \label{fig:system}
    \end{minipage}
    \hfill
    \begin{minipage}[b]{0.47\textwidth}
        \centering
        \includegraphics[width=\textwidth]{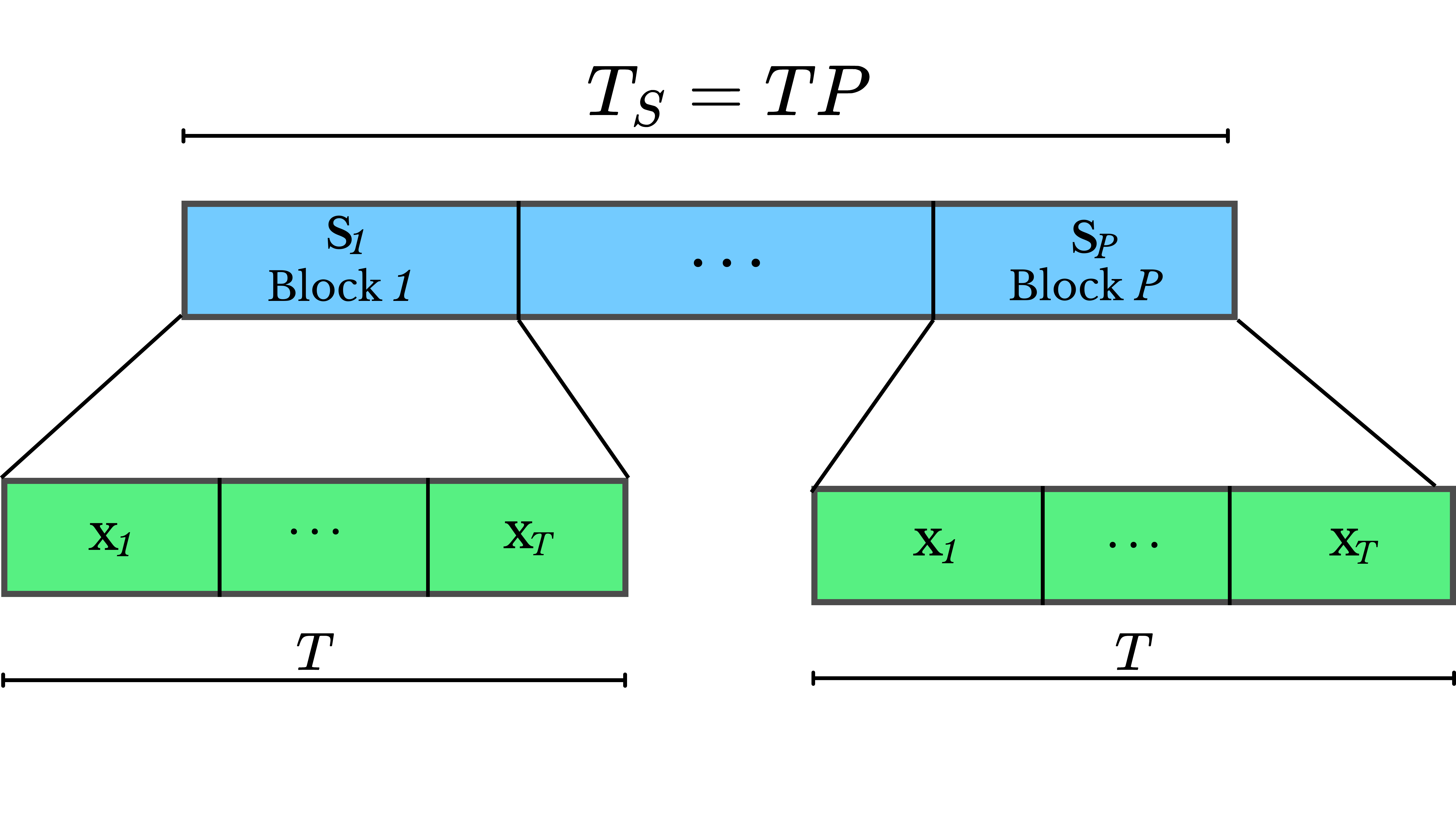}
        \caption{Transmission time structure.}
        \label{fig:time}
    \end{minipage}
    \label{fig:geral}
\end{figure}

\section{System model}
We consider the uplink of an \ac{MU} communication system where the base station (BS) receives data transmitted simultaneously by $K$ single-antenna users. The BS is equipped with an array of movable antennas composed of $N$ ports linked to $M$ antennas by a fast switching mechanism, with $M \leq N$. Each antenna is connected to a radio frequency (RF) chain and can connect to one of the $N$ available ports, as illustrated in Figure \ref{fig:system}. Following the idea of \cite{Zhang_2025}, port selection is performed from a binary switching matrix $\mathbf{S} \in \mathbb{R}^{M \times N}$, with $\mathbf{S}_{(m,n)} \in \{0,1\}$ having exactly only one unity element in each row and column indicating that the $m$-th antenna is (or not) connected at the $n$-th port, i.e.,
\begin{equation}
    \|\mathbf{S}_{(:,m)}\| = \|\mathbf{S}_{(n,:)}\| = 1,
    \label{EQ: S constraint}
\end{equation} 
such that $\mathbf{S} \mathbf{S}^{\mathrm{H}} = \mathbf{I}_M$. We assume block fading and a transmission time structure consisting of $P$ blocks, each comprising $T$ time slots, as depicted in Figure \ref{fig:time}. Each user 
sends a sequence of $T$ symbols within a block, while the switching matrix remains constant during a block but varies from block to block. The signal received by the BS at the $t$-th time slot in the $p$-th block is given by
\begin{equation}
\mathbf{y}_{p,t}  = \mathbf{S}_p\sum_{k = 1}^{K}\mathbf{h}_{k}x_{k,t} + \mathbf{z}_{p,t}, \label{EQ Received Signal SUM} 
\end{equation}  
where $\mathbf{h}_{k} \in \mathbb{C}^{N \times 1}$ and $x_{k,t}$ denote the vector channel and the transmitted data symbol associated with the $k$-th user, respectively, while
$\mathbf{z}_{p,t}$ is the additive white Gaussian noise (AWGN) vector term.

Alternatively, in a more compact form, the received signal (\ref{EQ Received Signal SUM}) can be rewritten as
\begin{equation}
\mathbf{y}_{p,t} = \mathbf{S}_p\mathbf{H}\mathbf{x}_{t} + \mathbf{z}_{p,t},
\label{EQ:Rx signal vec notation}
\end{equation}
where $\mathbf{H} = [\mathbf{h}_1, \mathbf{h}_2, \ldots, \mathbf{h}_K] \in \mathbb{C}^{N \times K}$ and $\mathbf{x}_t = [x_{1,t} \; x_{2,t} \; \dots \; x_{K,t}]^{\text{T}} \in \mathbb{C}^{K \times 1}$. To explore the multi-block structure of the proposed transmission protocol, the data symbols of each user are encoded across the $P$ blocks. Therefore, the received signal (\ref{EQ:Rx signal vec notation}) becomes
\begin{equation}
\mathbf{y}_{p,t} = \mathbf{S}_p\mathbf{H}\text{diag}(\mathbf{c}_p)\mathbf{x}_{t} + \mathbf{z}_{p,t}.
\label{EQ: Rx with coding}
\end{equation}

Finally, by collecting the received signal (\ref{EQ: Rx with coding}) across all the $t = 1, \ldots, T$ time slots, we get
\begin{equation}
    \mathbf{Y}_{p} = \mathbf{S}_p\mathbf{H}\text{diag}(\mathbf{c}_p)\mathbf{X} + \mathbf{Z}_{p}, 
\end{equation}
or, equivalently,
\begin{equation}
    \mathbf{Y}_{p} = \mathbf{S}_p\mathbf{H}D_p(\mathbf{C})\mathbf{X} + \mathbf{Z}_{p},
    \label{EQ: Received Signal}
\end{equation}
where $\mathbf{Y}_{p} = [\mathbf{y}_{p,1}, \mathbf{y}_{p,2}, \ldots, \mathbf{y}_{p,T}] \in \mathbb{C}^{M \times T}$, $\mathbf{X} = [\mathbf{x}_{1}, \mathbf{x}_{2}, \ldots, \mathbf{x}_{T}] \in \mathbb{C}^{K \times T}$ contains the data symbols transmitted by the $K$ users during $T$ time slots, $\mathbf{Z}_{p} = [\mathbf{z}_{p,1}, \mathbf{z}_{p,2}, \ldots, \mathbf{z}_{p,T}] \in \mathbb{C}^{M \times T}$ denotes the AWGN noise matrix, while $\mathbf{C} = [\mathbf{c}_{1}, \mathbf{c}_{2}, \ldots, \mathbf{c}_{P}]^{\text{T}} \in \mathbb{C}^{P \times K}$ is the block coding matrix whose each row contains the set of coding coefficients used by the $K$ users at the $p$-th block.


Note that the received signal in (\ref{EQ: Received Signal}) can be interpreted as the $p$-th, $p = 1, \ldots, P$, frontal slice of the third-order tensor $\mathcal{Y} \in \mathbb{C}^{M \times P \times T}$ that follows a PARAFAC2 model \cite{parafac2}. By analogy with (\ref{EQ: generic PARAFAC model}), the following correspondences can be established: 
\begin{equation}
    \big(\mathbf{A}_{p} \; , \mathbf{B} \; , \mathbf{C}\big) \leftrightarrow \big(\mathbf{S}_{p}\mathbf{H} \; , \mathbf{X}^{\text{T}} \; , \mathbf{C}\big).
\end{equation}

In the following, by exploiting the multilinear structure of the received signal (\ref{EQ: Received Signal}), we develop the PARAFAC2-based BALS semi-blind receiver to solve the joint channel and symbol estimation problem in communication systems with movable antennas. 



\section{Proposed BALS Semi-Blind Receiver}
To jointly estimate the channel users and the transmitted data symbols, we solve the following optimization problem:
\begin{equation}
    \Big(\hat{\mathbf{H}}, \; \hat{\mathbf{X}}\Big)  = \underset{\mathbf{H}, \mathbf{X}}{\arg\min} \,\, \left\|\mathbf{Y}_p - \mathbf{S}_p \mathbf{H} \mathbf{D}_p(\mathbf{C}) \mathbf{X}\right\|_\text{F}^2.
    \label{Eq:nonlinear problem}
\end{equation}

Since (\ref{Eq:nonlinear problem}) is a nonlinear problem, we propose estimating $\mathbf{H}$ and $\mathbf{X}$ using two separate linear subproblems that can be easily solved in the least squares (LS) sense by resorting to the well-known alternating least squares algorithm  \cite{comon_2009}. The main steps of the proposed solution are detailed below. 

\subsection{Channel Estimation} From (\ref{EQ: Received Signal}), we obtain the vectorized version of the received signal, which is given by
\begin{equation}
\begin{aligned}
    \text{vec}(\mathbf{Y}_p) & = \text{vec}(\mathbf{S}_p \mathbf{H} \mathbf{D}_p(\mathbf{C}) \mathbf{X})\\
     & =  \left( \mathbf{X}^{\text{T}}\mathbf{D}_p(\mathbf{C}) \otimes \mathbf{S}_p \right) \text{vec}\left(\mathbf{H}\right).
\end{aligned} 
\end{equation}
Applying the mixed-product property of the Kronecker product in (\ref{Eq:Propertie Mixed}), we obtain
\begin{equation}
    \mathbf{y}_p  = (\mathbf{X}^{\text{T}} \otimes \mathbf{I}_M) (\mathbf{D}_p(\mathbf{C}) \otimes \mathbf{S}_p) \, \mathbf{h}\;,
\end{equation}
where $\mathbf{y}_p = \text{vec}(\mathbf{Y}_p)$ and $\mathbf{h} = \text{vec}(\mathbf{H})$. Defining $\mathbf{F}_p = \mathbf{D}_p(\mathbf{C}) \otimes \mathbf{S}_p$, we can rewrite $\mathbf{y}_p$ as
\begin{equation}
    \mathbf{y}_p = (\mathbf{X}^{\text{T}} \otimes \mathbf{I}_M)\mathbf{F}_p\mathbf{h}.
    \label{EQ: Vectorized form of Yp}
\end{equation}
Stacking the observations across all the $p = 1, \ldots, P$ time slots, the complete system becomes
\begin{equation}
    \mathbf{y} =
    \begin{bmatrix}
        \text{vec}(\mathbf{Y}_1) \\
        \text{vec}(\mathbf{Y}_2) \\
        \vdots \\
        \text{vec}(\mathbf{Y}_P)
    \end{bmatrix}
    =
    \begin{bmatrix}
        \left( \mathbf{X}^{\text{T}} \otimes \mathbf{I}_M \right) \mathbf{F}_1 \\
        \left( \mathbf{X}^{\text{T}} \otimes \mathbf{I}_M \right) \mathbf{F}_2 \\
        \vdots \\
        \left( \mathbf{X}^{\text{T}} \otimes \mathbf{I}_M \right) \mathbf{F}_P
    \end{bmatrix}
    \mathbf{h},
\end{equation}
or equivalently,
\begin{equation}
\mathbf{y} = \mathbf{W}\mathbf{h},
\end{equation}
where 
\begin{equation}
    \mathbf{W} = \begin{bmatrix}
        \left( \mathbf{X}^{\text{T}} \otimes \mathbf{I}_M \right) \mathbf{F}_1 \\
        \left( \mathbf{X}^{\text{T}} \otimes \mathbf{I}_M \right) \mathbf{F}_2 \\
        \vdots \\
        \left( \mathbf{X}^{\text{T}} \otimes \mathbf{I}_M \right) \mathbf{F}_P
    \end{bmatrix} \in \mathbb{C}^{TMP \times NK}.
\end{equation}
Therefore, the estimate to the channel users can be directly obtained by solving the following LS problem:
\begin{equation}
    \hat{\mathbf{h}} =  \underset{\mathbf{h}}{\arg\min}\|\mathbf{y} - \mathbf{W}\mathbf{h}\|_\text{F}^2,
    \label{EQ: OPT problem to H}
\end{equation}
the solution of which is given by
\begin{equation}
    \hat{\mathbf{h}}  = \mathbf{W}^\dagger\mathbf{y} \quad \text{and} \quad \hat{\mathbf{H}} = \text{unvec}_{N \times K}(\hat{\mathbf{h}}).
    \label{EQ: Estimate H}
\end{equation}

\subsection{Data Estimation}
Additionally, to estimate the MU symbol matrix $\mathbf{X}$, we solve the following LS problem: 
\begin{equation}
    \hat{\mathbf{X}} =  \underset{\mathbf{X}}{\arg\min}\|\mathbf{Y} - \mathbf{Z}\mathbf{X}\|_\text{F}^2,
    \label{EQ: OPT problem to X}
\end{equation}
whose solution is given by
\begin{equation}
    \hat{\mathbf{X}} = \mathbf{Z}^\dagger\mathbf{Y},
    \label{EQ:Estimate X}
\end{equation}
where 
    \begin{equation}
    \mathbf{Y} =
    \begin{bmatrix}
        \mathbf{S}_1 \mathbf{H} \mathbf{D}_1(\mathbf{C}) \\
        \mathbf{S}_2 \mathbf{H} \mathbf{D}_2(\mathbf{C}) \\
        \vdots \\
        \mathbf{S}_P \mathbf{H} \mathbf{D}_P(\mathbf{C})
    \end{bmatrix}
    \mathbf{X} \in \mathbb{C}^{PM \times T},
\end{equation}
and
\begin{equation}
    \mathbf{Z} =   \begin{bmatrix}
        \mathbf{S}_1 \mathbf{H} \mathbf{D}_1(\mathbf{C}) \\
        \mathbf{S}_2 \mathbf{H} \mathbf{D}_2(\mathbf{C}) \\
        \vdots \\
        \mathbf{S}_P \mathbf{H} \mathbf{D}_P(\mathbf{C})
    \end{bmatrix} \in \mathbb{C}^{PM \times K}.
    \label{EQ: Matrix Z}
\end{equation}

From (\ref{EQ: Estimate H}) and (\ref{EQ:Estimate X}), the estimates of $\hat{\mathbf{H}}$ and $\hat{\mathbf{X}}$ can be obtained through the iterative bilinear alternating least squares (BALS) procedure which alternatively update one of the factor matrices to minimize the data fitting error while keeping the other one fixed. The pseudocode for our proposed solution is summarized in \textbf{Algorithm 1}. 
In the next section, we discuss the identifiability issues and computational complexity involved in our solution.
\begin{algorithm}
\caption{BALS semi-blind receiver}
\label{alg:BALS}
\KwIn{Initialize $\hat{\mathbf{X}}_{(i=0)}$ randomly and set $i = 0$}
\KwOut{Estimated $\hat{\mathbf{H}}$, $\hat{\mathbf{X}}$}
\Begin{
    \While{$\|\epsilon(i) - \epsilon(i-1)\| \geq \delta$}{
        1. Construct the sensing matrix $\mathbf{W}$ from $\hat{\mathbf{X}}_{(i)}$, coding matrix $\mathbf{C}$, and antenna switching matrices $\mathbf{S}_p$, $p = 1, \ldots, P$;
        
        2. Estimate the channel vector $\hat{\mathbf{h}}$ by solving the least-squares problem
        \begin{equation*}
            \hat{\mathbf{h}} = \mathbf{W}^{\dagger} \mathbf{y}.
        \end{equation*}

        3. Estimate the symbol matrix $\hat{\mathbf{X}}$ by solving the least-squares problem
        \begin{equation*}
            \hat{\mathbf{X}} = \mathbf{Z}^{\dagger} \mathbf{Y}.
        \end{equation*}
        
        4. Update iteration counter: $i \leftarrow i + 1$\;
        \textbf{end}
    }\textbf{end}
}
\end{algorithm}

\section{Indentifiability and Complexity}
Note that the uniqueness in the LS sense of this joint channel and symbol estimation algorithm from (\ref{EQ: Estimate H}) and (\ref{EQ:Estimate X}) requires that $TMP \geq NK$ and $PM \geq K$, respectively. These inequalities provide applicable constraints for the system parameters, such as the required number of switching blocks, the maximum number of users, the number of movable antennas, and the number of ports. For example, $K \leq \min\left(\dfrac{TMP}{N}, PM\right)$ guides us to a suitable choice for the values of the system parameters ensuring uniqueness of the estimated parameters.


Regarding computational complexity, the cost per iteration of \textbf{Algorithm 1} is dominated (in each step) by the SVD calculation in the pseudoinverse operations in steps $(2)$ and $(3)$. Let us recall that computing the SVD of a matrix $P \times Q$ has a complexity cost of $\mathcal{O}(PQ\min(P,Q))$. Thus, the complexity per iteration of our proposed solution is approximately given by $\mathcal{O}\big(PTM (NK)^2 + PMK^2\big)$, indicating that the complexity increases linearly with the number of antennas and blocks, and quadratically with the number of users and ports.

\begin{figure}[!t]
    \centering
    \begin{minipage}{0.55\textwidth}
        \centering
        \includegraphics[width=.85\textwidth]{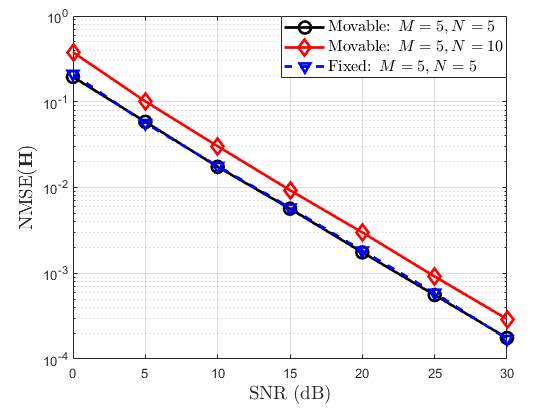}
        \caption{NMSE vs. SNR (dB)}
        \label{fig:NMSE}
    \end{minipage}
    \hfill
    \begin{minipage}{0.55\textwidth}
        \centering
        \includegraphics[width=.85\textwidth]{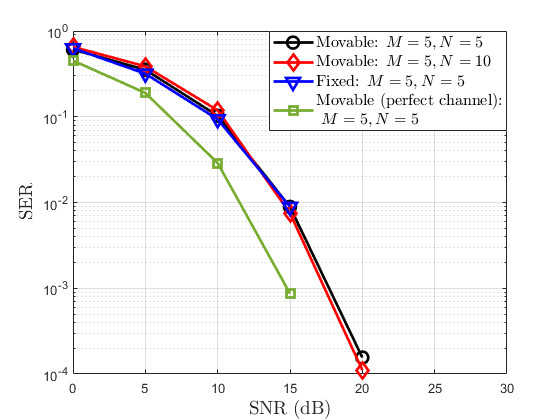}
        \caption{SER vs. SNR (dB)}
        \label{fig:SER}
    \end{minipage}
      \label{fig:geral}
\end{figure}

\section{Simulation Results}
Figures \ref{fig:NMSE} and \ref{fig:SER} show some performance results of the proposed movable-antenna-based semi-blind receiver. The results represent an average of 1000 independent Monte Carlo runs. The transmitted symbols are drawn from a 16-PSK constellation. The block coding matrix $\mathbf{C}$ is a $P\times K$ truncated DFT matrix. Our simulation setup assumes $M=5$ movable antennas, $K=5$ users, $P=8$ blocks, and $T=10$ time slots. The number of ports is set to $N=5$ or $N=10$. As a reference for comparison, we plot the performance of the baseline fixed antenna case, where the antenna positions do not change (which corresponds to having $\mathbf{S}_p=\mathbf{I}_N$, $p=1, \ldots, P$). Figure \ref{fig:NMSE} depicts the channel estimation performance regarding the normalized mean squared error (NMSE). In this scenario, we can observe that the accuracy of the estimated \ac{MU} channel $\mathbf{H}$ decreases when the number of ports increases. This result is expected since increasing the possible antenna positions implies estimating more channel coefficients.
On the other hand, more spatial degrees of freedom are available for optimizing the transmit beamforming in the downlink transmission stage. Spectral efficiency results will be shown in the full version of the paper. Figure \ref{fig:SER} shows the average symbol error rate (SER) over all the users. The results show that the proposed semi-blind receiver can decode the transmitted data effectively. At the same time, the recovery of the data symbols is insensitive to the number of ports in this specific setup.

\begin{figure}[!t]
    \begin{minipage}{0.55\textwidth}
        \centering
        \includegraphics[width=.85\textwidth]{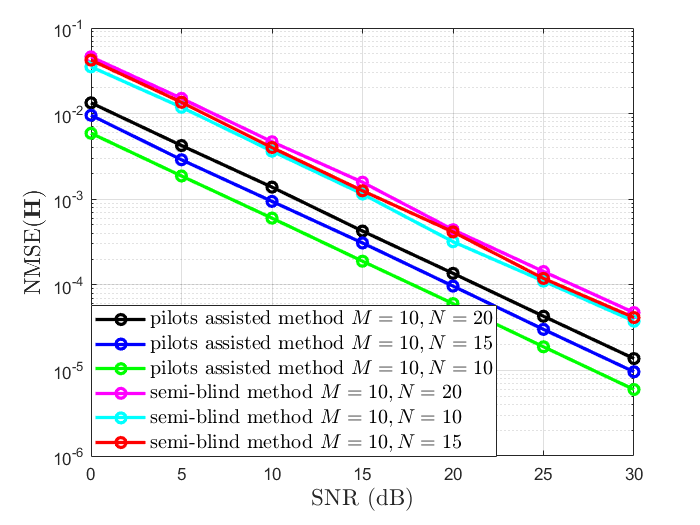}
        \caption{NMSE vs. SNR (dB). } 
        \label{fig:pilot}
    \end{minipage}
      \begin{minipage}{0.55\textwidth}
        \centering
        \includegraphics[width=.85\textwidth]{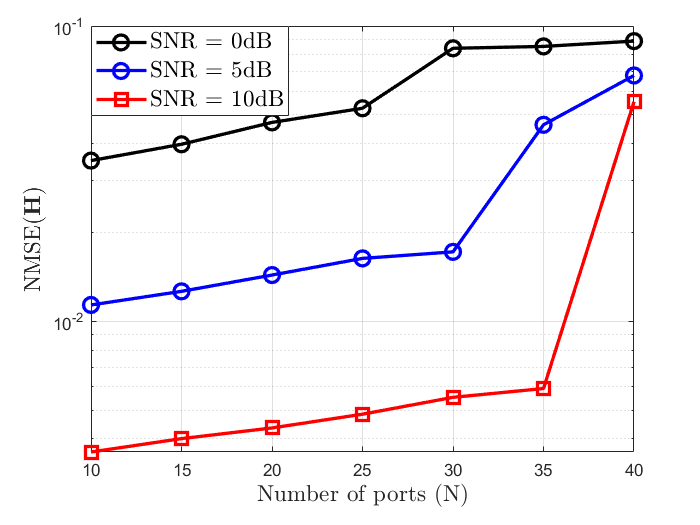}
        \caption{NMSE vs. Number of ports ($N$).}
          \label{fig:Nport}
    \end{minipage}
      \label{fig:geral2}
\end{figure}

Figure \ref{fig:pilot} evaluates the performance of the proposed receiver with a reference pilot-assisted approach in which the channel estimate is obtained using the least squares method based on a known pilot matrix $\mathbf{X}$ directly from (\ref{EQ: Estimate H}). From this figure, as expected, the pilot-assisted channel estimate is more accurate than the proposed semi-blind receiver. 
On the other hand, our approach allows earlier data decoding during the channel estimation stage, providing joint estimates of the channel and data for all users, which can increase the system spectral efficiency in this stage.

In Figure \ref{fig:Nport}, the channel estimation performance is examined as a function of the number of available ports $N$. Clearly, the performance worsens when the number of ports increases. In particular, the estimation accuracy reaches its upper limit (worst case) for configurations close to $N = 40$ ports, showing very similar results regardless of the SNR. These results are in agreement with those of Figure \ref{fig:NMSE}, leading to similar conclusions, i.e., increasing the number of ports implies estimating more channel coefficients, which leads to performance degradation.


\section{Conclusions}
This paper presented a semi-blind receiver for joint channel and symbol estimation in uplink \ac{MU} communication systems equipped with movable antennas. By exploiting the multilinear structure of the received signals and adopting a PARAFAC2 tensor decomposition, the proposed approach enables estimation without relying on prior training sequences. Simulation results demonstrate that the method maintains low symbol error rates even when the number of ports increases, although the accuracy of the channel estimation degrades due to the higher number of coefficients to be estimated. Nevertheless, the system remains effective at recovering the transmitted data. The ability to explore more spatial degrees of freedom with reduced hardware complexity makes this architecture a promising alternative to conventional fixed-antenna systems. 

Further work may consider extending this framework to time-varying channels while optimizing the downlink transmission based on the estimated CSI. Another perspective includes exploiting the sparse structure of the channel linking the UEs to the BS. The extension to a scenario with multi-antenna UEs is also of interest and will be addressed in an extended version of this paper.



\bibliographystyle{IEEEtran}
\bibliography{IEEEexample}

\end{document}